\documentclass[preprint,12pt]{elsarticle}
\usepackage[dvipdfm,
            pdfstartview=FitH,
            CJKbookmarks=true,
            bookmarksnumbered=true,
            bookmarksopen=true,
           colorlinks,
            pdfborder=001,
            linkcolor=black,
            anchorcolor=black,
            citecolor=black
            ]{hyperref}

\usepackage[fleqn]{amsmath}

\usepackage{amsmath}
\usepackage{extarrows}
\usepackage{multirow}
\usepackage{makecell}
\usepackage{subfigure}
\usepackage{bm}

\usepackage{graphicx}
\usepackage{array}
\usepackage{booktabs}
\usepackage{caption}
\usepackage[nodots]{numcompress}
\usepackage{array}
\usepackage{listings}
\usepackage{multirow}
\usepackage{listings}
\usepackage{extarrows}
\usepackage{amsmath}
\usepackage{amsfonts,amssymb}
\usepackage{makecell}
\usepackage{bm}

\usepackage{graphicx}
\usepackage{array}
\newcommand{\PreserveBackslash}[1]{\let\temp=\\#1\let\\=\temp}
\newcolumntype{C}[1]{>{\PreserveBackslash\centering}p{#1}}
\newcolumntype{R}[1]{>{\PreserveBackslash\raggedleft}p{#1}}
\newcolumntype{L}[1]{>{\PreserveBackslash\raggedright}p{#1}}

\journal{}

\begin{document}

\begin{frontmatter}
\title{Calculating correlation coefficient for  Gaussian copula }

\author{Qing Xiao\corref{cor}}
\ead{xaoshaoying@shu.edu.cn}

%
\cortext[cor]{Corresponding author}
\begin{abstract}
When Gaussian copula with linear correlation coefficient is used to model correlated   random variables, one crucial issue is to determine a suitable correlation coefficient $\rho_z$ in normal space for two   variables with correlation coefficient $\rho_x$. This paper  attempts to address this problem. For two continuous variables, the marginal transformation  is approximated by a weighted sum of Hermite polynomials, then, with Mehler's   formula,  a polynomial of $\rho_z$ is derived to approximate the function relationship between $\rho_x$ and $\rho_z$. If a discrete variable is involved, the marginal transformation is decomposed into piecewise continuous ones, and $\rho_x$ is  expressed as a polynomial of $\rho_z$ by Taylor expansion. For a given $\rho_x$, $\rho_z$ can be efficiently determined by solving a polynomial equation.

\end{abstract}

\end{frontmatter}

\section{Introduction}
Gaussian copula has been widely used to model correlated non-normal vector  $\bm{X}=(x_{1},\dots,x_{i},\dots,x_{m})^T$
\cite{NatafC,NatafC2}. With the marginal transformation of copula, a non-normal variable  $x$  can be mapped to the standard normal space:
\begin{equation}\label{CopulaD}
z=\Phi^{-1}[F(x)],
\end{equation}
where $z$ is a standard normal variable, $\Phi^{-1}(\cdot)$ is the inverse cumulative distribution function (CDF) of $z$. $F(\cdot)$ is the CDF of $x$. 

For a correlated random vector, it requires to determine a suitable correlation matrix $\bm{R_Z}$ in normal space to well represent the dependency structure of $\bm{X}$. That's to say, for each correlation coefficient $\rho_x(i,j)$ ($i\neq j$) between $x_i$ and $x_j$, an appropriate value of $\rho_z(i,j)$ in $\bm{R_Z}$ should be determined.

Rewrite Eq.\eqref{CopulaD} in an inverse form:
\begin{equation}\label{CopulaDxs2}
x=F^{-1}[\Phi(z)],
\end{equation}
where  $\Phi(\cdot)$ is the CDF of $z$. $F^{-1}(\cdot)$ is the inverse CDF of $x$. Then, for a given $\rho_x$ between $x_i$ and $x_j$, it has the following relationship with $\rho_z$:
\begin{equation}\label{NatafEq1}
\begin{split}
  \rho_x\sigma_i\sigma_j&+\mu_i\mu_j = E[x_ix_j]= \int_{-\infty}^{+\infty}\int_{-\infty}^{+\infty}F^{-1}_{i}[\Phi(z_i)]F^{-1}_{j}[\Phi(z_j)]\phi(z_i,z_j,\rho_z) dz_idz_j,
\end{split}
\end{equation}
where $\mu_i$, $\mu_j$ denote the means of $x_i$, $x_j$, respectively, $\sigma_i$, $\sigma_j$ denote the standard deviations respectively.  $\phi(z_i,z_j,\rho_z)$ is the joint PDF of two correlated standard normal variables:
\begin{equation}\label{doublenormal}
\phi(z_i,z_j,\rho_z)=\frac{1}{2\pi\sqrt{1-\rho_z^2}}e^{-\frac{z_i^2-2\rho_zz_iz_j+z_j^2}{2(1-\rho_z^2)}}.
\end{equation}

In most cases, the integral equation in Eq.\eqref{NatafEq1} is difficult to be solved analytically, and numerical methods should be employed to determine $\rho_z$. If $x_i$ and $x_j$ are both continuous random variables,  49 empirical formulae have been derived to calculate $\rho_z$\cite{DKL1}; three empirical formulae based on Johnson system are also developed\cite{Johnson}. Furthermore, because $\rho_x$ is a continuous function of $\rho_z$, which is located in the interval $[-1,1]$, a root finding method can be used to estimate $\rho_z$ for a given $\rho_x$ \cite{NORTA2,Nataf4}.   If   $x_i$ and $x_j$  are both discrete random variables,   another root finding algorithm is also developed  to determine $\rho_z$\cite{NatafDDK} .

Except for the  empirical formulae,  other methodologies are inconvenient to establish the function relationship between  $\rho_z$ and $\rho_x$, an issue this paper attempts to address.   The basic idea is to employ a polynomial of $\rho_z$ to approximate the function relationship between $\rho_z$ and $\rho_x$.  With Hermite polynomials and Mehler's   formula,  all three possible scenarios: continuous case, discrete case and mixed case are considered.  For a given $\rho_x$, $\rho_z$ can be efficiently calculated by solving a polynomial equation.



\section{Continuous case}\label{CoefficientContinuous}


If both $x_i$ and $x_j$ are continuous  variables, represent the transformation in Eq.\eqref{CopulaDxs2} by:
\begin{equation}\label{Polynomial}
x=F^{-1}\left[\Phi(z)\right] = \sum_{k=0}^{\infty}a_{k}H_{k}(z),
\end{equation}
where $a_{k}$ ($k=0,1,\dots$) are undetermined coefficients. $H_{k}(z)$ is  the $k$th-order Hermite polynomial, which is defined by\cite{Hermite}:
\begin{equation}\label{cwe2}
  H_{k+1}(z)=zH_k(z)-H_k^{'}(z),~~  H_1(z)=z,~~H_0(z)=1.
\end{equation}
Hermite polynomial  has the following property:
\begin{equation}\label{Hermite}
\int_{-\infty}^{+\infty}H_{m}(z)H_{k}(z)\phi(z)dz=\Big \{
             \begin{array}{ll}
               k! ~~~~~ m=k\\
               0~~~~~~m\neq k.
             \end{array}
\end{equation}

Using this property, $a_{k}$  can be  easily determined. Consider the following equation:
\begin{equation}\label{CoeffA}
\begin{split}
\int_{-\infty}^{+\infty}H_{m}(z)F^{-1}\left[\Phi(z)\right] \phi(z)dz=&\int_{-\infty}^{+\infty}H_{m}(z)\sum_{k=0}^{\infty}a_{k}H_{k}(z)\phi(z)dz\\
=&\sum_{k=0}^{\infty}a_{k}\int_{-\infty}^{+\infty}H_{m}(z)H_{k}(z)\phi(z)dz
\end{split}
\end{equation}
If one needs to determine $a_k$, set $m=k$, and Eq.\eqref{CoeffA} becomes:
\begin{equation}
\begin{split}
\int_{-\infty}^{+\infty}H_{k}(z)F^{-1}\left[\Phi(z)\right]\phi(z)dz=a_{k}\cdot k!,
\end{split}
\end{equation}
then
\begin{equation}\label{vcc}
\begin{split}
a_k=\frac{1}{k!}\int_{-\infty}^{+\infty}H_k(z)F^{-1}[\Phi(z)] \phi(z)dz.
\end{split}
\end{equation}
The above integral can be accurately calculated by an $m$-point Gauss-Hermite quadrature ($m>k$).

Using  Hermite polynomials defined by Eq.\eqref{cwe2}, the Mehler's formula can be expressed as\cite{Mehler}:
\begin{equation}\label{m}
\frac{1}{\sqrt{1-\rho_z^2}}exp\left({-\frac{\rho_z^2(z_i^2+z_j^2)-2\rho_zz_iz_j}{2(1-\rho_z^2)}}\right)=\sum_{k=0}^{\infty}H_k(z_i)H_k(z_j)\frac{\rho_z^k}{k!}.
\end{equation}
Then, $\phi(z_i,z_j,\rho_z)$ in Eq.\eqref{doublenormal} can be expressed as:
\begin{equation}\label{doublenormal2}
\phi(z_i,z_j,\rho_z)=\phi(z_i)\phi(z_j) \sum_{k=0}^{\infty}H_k(z_i)H_k(z_j)\frac{\rho_z^k}{k!}.
\end{equation}

 Let $x_i$ and $x_j$    be approximated by an $n$th-order polynomial of $z_i$ and $z_j$ respectively:
\begin{equation}\label{Polynomial22}
\begin{split}
x_i &\simeq \sum_{k_i=0}^{n}a_{i,k_i}H_{k_i}(z_i),~~x_j\simeq \sum_{k_j=0}^{n}a_{j,k_j}H_{k_j}(z_j).
\end{split}
\end{equation}
Substitute  Eq.\eqref{doublenormal2} and Eq.\eqref{Polynomial22} into Eq.\eqref{NatafEq1}:
\begin{equation}\label{DDD}
  \begin{split}
     E[x_ix_j] \simeq  &\int_{-\infty}^{+\infty}\int_{-\infty}^{+\infty} \sum_{k_i=0}^{n}a_{i,k_i}H_{k_i}(z_i)  \cdot \sum_{k_j=0}^{n}a_{j,k_j}H_{k_j}(z_j)\cdot \phi(z_i)\phi(z_j) \sum_{k=0}^{\infty}\frac{\rho_z^k}{k!}H_k(z_i)H_k(z_j)dz_idz_j\\
    = &\sum_{k=0}^{\infty} \frac{\rho_z^k}{k!} \sum_{k_i=0}^{n}\sum_{k_j=0}^{n}a_{i,k_i}\int_{-\infty}^{+\infty}H_{k_i}(z_i) H_k(z_i)\phi(z_i) dz_i \cdot a_{j,k_j}\int_{-\infty}^{+\infty}  H_{k_j}(z_j)H_k(z_j) \phi(z_j)dz_j.\\
  \end{split}
\end{equation}
According to Eq.\eqref{Hermite},  the coefficient of $\rho_z^k$ is not zero only if $k_i=k_j=k$, and it has:
\begin{equation}\label{wwww}
\rho_x\sigma_i\sigma_j+\mu_i\mu_j=E[x_ix_j] \simeq \sum_{k=0}^{n}k!a_{i,k}a_{j,k}\rho_z^k.
\end{equation}


As shown in Eq.\eqref{wwww}, $\rho_x$ is expressed as an $n$th-order polynomial of $\rho_z$.   For  a given  $\rho_x$ between $x_i$ and $x_j$, calculate the coefficients $a_{i,k}$ and $a_{j,k}$ by Eq.\eqref{vcc},  then, solve the polynomial equation in Eq.\eqref{wwww},  the valid value of $\rho_z$  is restricted by:
\begin{equation}\label{Condition}
  -1<\rho_z<1~~\mbox{and}~~\rho_z\rho_x>0.
\end{equation}
\section{Discrete case}
This section develops a method to calculate $\rho_z$ for two discrete variables. Suppose the support of $x_i$ is $[X_{i,1},X_{i,2},\dots,X_{i,k_i},\dots,X_{i,N_i}]$. Denote:
\begin{equation}\label{CopulaD2}
Z_{i,k_i}=\Phi^{-1}[F_i(X_{i,k_i})],~~k_i=0,1,\dots,N_i.
\end{equation}
where $Z_{i,0}=\Phi^{-1}[F_i(X_{i,0})]=-\infty$. Then:
\begin{equation}\label{CopulaD32}
\mbox{for}~~Z_{i,k_{i}-1} < z_i\leq Z_{i,k_{i}},~~x_i=F_i^{-1}[\Phi(z_i)]=X_{i,k_i},~~k_i=1,\dots,N_i.
\end{equation}

Using Eq.\eqref{CopulaD32},   Eq.\eqref{NatafEq1} can be expressed as:
\begin{equation}\label{finall-1}
\begin{split}
  \rho_x=&-\frac{\mu_i\mu_j}{\sigma_i\sigma_j}+\frac{1}{\sigma_i\sigma_j}\sum_{k_i=1}^{N_i}\sum_{k_j=1}^{N_j}X_{i,k_i}X_{j,k_j}\int_{Z_{i,{k_{i}-1}}}^{Z_{i,{k_{i}}}}\int_{Z_{j,{k_{j}-1}}}^{Z_{j,{k_{j}}}}\phi(z_i,z_j,\rho_z) dz_idz_j.
\end{split}
\end{equation}
The double integral in   Eq.\eqref{finall-1} is performed over a rectangular region: $D: {Z_{i,{k_{i}-1}}}\leq z_i \leq{Z_{i,{k_{i}}}}$, ${Z_{j,{k_{j}-1}}}\leq z_j \leq{Z_{j,{k_{j}}}}$.  By Green's theorem, this double integral can be transformed into curvilinear integral:
\begin{equation}\label{finall-2}
\begin{split}
  \rho_x =&-\frac{\mu_i\mu_j}{\sigma_i\sigma_j}+\frac{1}{\sigma_i\sigma_j}\sum_{k_i=1}^{N_i}\sum_{k_j=1}^{N_j}X_{i,k_i}X_{j,k_j}\big[\Phi(Z_{i,{k_{i}}},Z_{j,{k_{j}}},\rho_z)+\Phi(Z_{i,{k_{i}-1}},Z_{j,{k_{j}-1}},\rho_z)-\\
  &\Phi(Z_{i,{k_{i}-1}},Z_{j,{k_{j}}},\rho_z)-\Phi(Z_{i,{k_{i}}},Z_{j,{k_{j}-1}},\rho_z)\big],
\end{split}
\end{equation}
where $\Phi(z_i,z_j,\rho_z)$ is the joint CDF of two correlated standard normal variables.



Denote the function relationship between $\rho_z$ and $\rho_x$ as:
\begin{equation}
\rho_x=G(\rho_z).
\end{equation}
Take $n$th-order derivative on both sides of Eq.\eqref{finall-2}:
\begin{equation} \label{DDDD02}
\begin{split}
G^{(n)}(\rho_z) =&\frac{1}{  \sigma_i\sigma_j}\sum_{k_i=1}^{N_i}\sum_{k_j=1}^{N_j}X_{i,k_i}X_{j,k_j}\big[\phi^{(n-1)}(Z_{i,{k_{i}}},Z_{j,{k_{j}}},\rho_z)+\phi^{(n-1)}(Z_{i,{k_{i}-1}},Z_{j,{k_{j}-1}},\rho_z)-\\
  &\phi^{(n-1)}(Z_{i,{k_{i}}},Z_{j,{k_{j}-1}},\rho_z)-\phi^{(n-1)}(Z_{i,{k_{i}-1}},Z_{j,{k_{j}}},\rho_z)\big].
\end{split}
\end{equation}
Consider the Taylor expansion of $G(\rho_z)$:
\begin{equation}\label{Taylor}
\rho_x=G(0)+\frac{G^{'}(0)}{1!}\rho_z+\cdots+\frac{G^{(n)}(0)}{n!}\rho_z^n+\cdots,~~~G(0)=0.
\end{equation}
Because Taylor expansion of $\phi(z_{i},z_{j},\rho_z)$ at $\rho_z=0$ is:
\begin{equation}
\phi(z_{i},z_{j},\rho_z)=\sum_{k=0}^{\infty}\phi^{(k)}(z_{i},z_{j},0)\frac{\rho_z^k}{k!},\nonumber
\end{equation}
 according to Eq.\eqref{doublenormal2}, it has:
\begin{equation}\label{dsdv2222s}
\phi^{(n-1)}(z_{i},z_{j},0)=H_{n-1}(z_i)H_{n-1}(z_j)\phi(z_i)\phi(z_j).
\end{equation}


For two discrete random variables $x_i$ and $x_j$, the values of $Z_{i,k_{i}-1}$, $Z_{j,k_{j}-1}$, $Z_{i,k_i}$ and $Z_{j,k_j}$ can be obtained by Eq.\eqref{CopulaD2}, then, set $\rho_z=0$ in Eq.\eqref{DDDD02}, with Eq.\eqref{dsdv2222s},    $G^{(n)}(0)$ can be easily calculated, and   the coefficients of Taylor series in Eq.\eqref{Taylor} can be determined.  For a given $\rho_x$, solving the polynomial equation in Eq.\eqref{Taylor} gives the value of $\rho_z$,  the valid solution is restricted by Eq.\eqref{Condition}.

\section{Mixed case}
This section develops a method to calculate $\rho_z$ for a given $\rho_x$ between a discrete variable $x_i$ and a continuous variable $x_j$.  Suppose the support of $x_i$ is $[X_{i,1},\dots,X_{i,k_i},\dots,X_{i,N_i}]$. Using Eq.\eqref{CopulaD2}, Eq.\eqref{NatafEq1} can be rewritten as:
\begin{equation}\label{mix1}
\begin{split}
  \rho_x =-\frac{\mu_i\mu_j}{\sigma_i\sigma_j}+\frac{1}{\sigma_i\sigma_j}\sum_{k_i=1}^{N_i} X_{i,k_i}\int_{-\infty}^{+\infty} \int_{Z_{i,{k_{i}-1}}}^{Z_{i,{k_{i}}}}F^{-1}_j [\Phi(z_j)]\phi(z_i,z_j,\rho_z) dz_idz_j.
\end{split}
\end{equation}

As shown in Eq.\eqref{Taylor}, $\rho_x$ can be expressed as a polynomial of $\rho_z$,  the  problem is to calculate $G^{(n)}(\rho_z)|_{\rho_z=0}$. In Eq.\eqref{mix1}, $G^{(n)}(\rho_z)$ is:
\begin{equation}\label{mix2}
\begin{split}
G^{(n)}(\rho_z)=\frac{1}{\sigma_i\sigma_j}\sum_{k_i=1}^{N_i} X_{i,k_i} \int_{-\infty}^{+\infty}F^{-1}_j [\Phi(z_j)]\Bigg[\int_{Z_{i,{k_{i}-1}}}^{Z_{i,{k_{i}}}}\frac{\partial^n \phi(z_i,z_j,\rho_z)} {\partial\rho_z^n}  dz_i\Bigg]dz_j.
\end{split}
\end{equation}

Because:
\begin{equation}
 \frac{\partial\phi(z_i,z_j,\rho_z)} {\partial \rho_z}   =\frac{\partial^2 \phi(z_i,z_j,\rho_z)} { \partial z_i\partial z_j},
\end{equation}
then:
\begin{equation}\label{mix3sm}
\begin{split}
\int_{Z_{i,{k_{i}-1}}}^{Z_{i,{k_{i}}}} \frac{\partial^n \phi(z_i,z_j,\rho_z)} {\partial\rho_z^n}dz_i=&\int_{Z_{i,{k_{i}-1}}}^{Z_{i,{k_{i}}}} \frac{\partial^{n-1} } {\partial\rho_z^{n-1}}\left(\frac{\partial \phi(z_i,z_j,\rho_z)} {\partial \rho_z}\right)dz_i\\
=&\int_{Z_{i,{k_{i}-1}}}^{Z_{i,{k_{i}}}} \frac{\partial^{n} } {\partial\rho_z^{n-1}\partial z_j}\left(\frac{\partial\phi(z_i,z_j,\rho_z)}{\partial z_i}\right)  dz_i\\
=&\Bigg[\frac{\partial ^{n}\phi(z_{i,k_{i}-1},z_j,\rho_z)} {\partial\rho_z^{n-1}\partial z_j}\Bigg]_{Z_{i,k_{i}-1}}^{Z_{i,k_{i}}}.
\end{split}
\end{equation}
For Hermite polynomials, it holds that:
\begin{equation}
H_{k}(z)=zH_{k-1}(z)-H_{k-1}^{'}(z).
\end{equation}
Using this property and Eq.\eqref{dsdv2222s},  it can be derived that:
\begin{equation}\label{mix4}
\frac{\partial^{n} \phi(z_{i,k_{i}},z_j,0)}  {\partial\rho_z^{n-1}\partial z_j}=\frac{\partial } {\partial z_j}\left(\frac{\partial \phi^{n-1}(z_i,z_j,0)} {\partial \rho_z^{n-1}}\right)=-H_{n-1}(z_{i,k_i})\phi(z_{i,k_{i}})H_{n}(z_{j,k_j})\phi(z_{j,k_{j}}),
\end{equation}

In Eq.\eqref{mix3sm}, set $\rho_z=0$, using  Eq.\eqref{mix4}, it has:
\begin{equation} \label{mix3}
\int_{Z_{i,{k_{i}-1}}}^{Z_{i,{k_{i}}}}\frac{\partial^n \phi(z_i,z_j,0)} {\partial\rho_z^n}  dz_i=-H_{n}(z_{j})\phi(z_{j})\big[H_{n-1}(Z_{i,k_i})\phi(Z_{i,k_{i}})-
  H_{n-1}(Z_{i,k_{i}-1})\phi(Z_{i,k_{i}-1})\big].
\end{equation}


In Eq.\eqref{mix2}, set $\rho_z=0$, using Eq.\eqref{mix3}, it has:
\begin{equation}\label{mix5}
\begin{split}
 G^{(n)}(0) =&-\frac{1}{\sigma_i\sigma_j}\sum_{k_i=1}^{N_i} X_{i,k_i}\Big[H_{n-1}(Z_{i,k_i})\phi(Z_{i,k_{i}})-H_{n-1}(Z_{i,k_{i}-1})\phi(Z_{i,k_{i}-1})\Big]\cdot\\
  & \int_{-\infty}^{+\infty}F^{-1}_j [\Phi(z_j)]H_{n}(z_{j})\phi(z_{j})dz_j.
\end{split}
\end{equation}
For a discrete variable $x_i$ and a continuous variable $x_j$, calculate $G^{(n)}(0)$ ($n=1,2,\dots$) by Eq.\eqref{mix5} and substitute them into Eq.\eqref{Taylor}, then, a polynomial of $\rho_z$ can be obtained, which serves an approximation of $G(\rho_z)$.


\section{Determining the degree of polynomial}\label{DetermineDegree}
It should be noted that a closed form of $G(\cdot)$ can be obtained for several cases, which are shown in Appendix (although the results of \nameref{CASE1}, \nameref{CASE8} and \nameref{CASE9} are already widely known). For other cases, let the function relationship between $\rho_x$ and $\rho_z$ be approximated by an $n$th-order polynomial:
\begin{equation}
\rho_x=G(\rho_z)\simeq \sum_{k=0}^{n}b_k\rho_z^k,
\end{equation}

Because $\rho_z\in[-1,1]$, according to  Weierstrass approximation theorem\cite{saxe2002beginning}, $G(\cdot)$ can be approximated as closely as desired by a polynomial function of $\rho_z$. However,   Runge's theorem states that  a polynomial of too high degree would cause oscillation at the edges of the interval, which means going to higher degrees does not always improve accuracy\cite{Runge}. Therefore,    an appropriate degree of the polynomial should be chosen, such that $G(\cdot)$ can be well approximated. But this task may be difficult to  perform in a theoretical way. Here, an empirical method is  developed to determine the degree $n$.


Start from $j=1$, establish a $1$th-order polynomial by the proposed method,  then, increase the degree $j$ in a step of $\Delta n$,   obtain a polynomial sequence:
\begin{equation}
\begin{split}
P_1(\rho_z)=&\sum_{k=0}^{1}b_k\rho_z^k,\\
P_{1+\Delta n}(\rho_z)=&\sum_{k=0}^{1+\Delta n}b_k\rho_z^k\\
&\vdots\\
P_{j}(\rho_z)=&\sum_{k=0}^{j}b_k\rho_z^k\\
P_{j+\Delta n}(\rho_z)=&\sum_{k=0}^{j+\Delta n}b_k\rho_z^k\\
&\vdots
\end{split}\nonumber
\end{equation}
Choosing   a set values of $\rho_{z,i}$ on interval $[-1,1]$ in steps of $\Delta \rho_z$ (say, $\Delta \rho_z=0.01$, then $i=1,2,\dots,201$), then evaluate   the difference between two neighbouring polynomials, and select the maximum one:
\begin{equation}
\Delta P_j=max\{\left|{P_{j}(\rho_{z,i})-P_{j+\Delta n}(\rho_{z,i})}\right|\},~~i=1,2,\dots,201.
\end{equation}
$\Delta P_j$ denotes the maximum difference between a $j$th-order polynomial and a ($j+\Delta n$)th-order polynomial.

Set an small error bound $\delta$  for  $\Delta P_j$ (say $\delta=10^{-4}$), and a polynomial with a value of $\Delta P_j<\delta$ can be  expected to give a good approximation of $G(\rho_z)$. Suppose the optimal degree of the polynomial is $n$,  the underlying assumption is that as the degree $j$ ($j\leq n$) increases, the sequence would converge to an optimum polynomial, whose difference to the neighbouring polynomial should  not be significant. Here is an example to illustrate this method. 

Suppose $x_i$ and $x_j$ follow  Uniform distributions,  then (see Table \ref{gt}):
\begin{equation}
\rho_x=G(\rho_z)=\frac{6}{\pi}arcsin(\frac{\rho_z}{2}),
\end{equation}
Define:
\begin{equation}
\Delta P_j^*=max\left\{\left|{P_{j}(\rho_{z,i})-\frac{6}{\pi}arcsin(\frac{\rho_{z,i}}{2})}\right|\right\},~~i=1,2,\dots,201.
\end{equation}
$\Delta P_j^*$ denotes the difference between   a $j$th-order polynomial and the theoretical formula.

 \begin{table}[!htb]
\setlength{\abovecaptionskip}{0pt}
\centering
\caption{The values of $\Delta P_j$ and $\Delta P_j^*$ for $U(0,1) \thicksim U(0,1)$}
\label{optimaldegree}
{\begin{tabular}[l]{@{}ccccc}
\hline

Degree $j$&$\Delta P_j$&$\Delta P_j^*$\\
\hline
1	&	0.33	&	0.40	\\
3	&	0.054	&	0.065	\\
5	&	0.0091	&	0.011	\\
7	&	$1.6\times 10^{-3}$	&	$2.0\times 10^{-3}$	\\
9	&	$3.1\times 10^{-4}$	&	$3.9\times 10^{-4}$	\\
11	&	$6.2\times 10^{-5}$	&	$7.9\times 10^{-5}$	\\
13	&	$1.3\times 10^{-5}$	&	$1.6\times 10^{-5}$	\\
15	&	$2.7\times 10^{-6}$	&	$3.5\times 10^{-6}$	\\
17	&	$6.0\times 10^{-7}$	&	$7.7\times 10^{-7}$	\\

\hline
\end{tabular}}
\end{table}

Start from $1$th-order polynomial, increase the degree of the polynomial in steps of $\Delta n=2$, and establish  polynomials as described in Section \ref{CoefficientContinuous}, then, calculate $\Delta P_j$ and $\Delta P_j^*$. Several values are chosen and presented in Table \ref{optimaldegree}.

As can be seen,  the variation of $\Delta P_j$ agrees with the variation of $\Delta P_j^*$, and a $9$th-order polynomial can give a good approximation of $G(\rho_z)$. Testing for other eight  cases in Table \ref{gt}, this method stands a decent chance of finding a well-performing polynomial.


\section{Comparison with linear search method}
Rewrite Eq.\eqref{NatafEq1} in   following form:
\begin{equation}\label{NatafEq345}
\rho_x=-\frac{\mu_i\mu_j}{\sigma_i\sigma_j}+\frac{1}{\sigma_i\sigma_j}\int_{-\infty}^{+\infty}\int_{-\infty}^{+\infty}F^{-1}_{i}[\Phi(z_i)]F^{-1}_{j}[\Phi(z_j)]\phi(z_i,z_j,\rho_z) dz_idz_j,
\end{equation}
because the function relationship between $\rho_z$ and $\rho_x$ is continuous and strictly increasing\cite{ NatafDDK, NatafDDK2}, and $\rho_z$ is located in $[-1,1]$, for a given $\rho_x$, $\rho_z$ can also be determined through a linear search method.

Suppose it requires to determine $\rho_z$ for $\rho_x=\rho_x^*$, and a bisection method is employed to find the root of the integral equation in Eq.\eqref{NatafEq345}. If  the error bound of the result   is $\varepsilon$, it would need to evaluate the double integral $ T$ times at $T$ different values of $\rho_z$ ($T=\lceil 1-log_{2}\varepsilon\rceil$, if $\varepsilon=10^{-3}$, $T= 11$).
\subsection{Continuous case}\label{CompareContinuous}
If $x_i$ and $x_j$ are both continuous random variables, substitute $x_i=u_i$, $x_j=\rho_zu_i+\sqrt{1-\rho_z^2}u_j$ into Eq.\eqref{NatafEq345}:
\begin{equation} \label{ppp}
\rho_x=-\frac{\mu_i\mu_j}{\sigma_i\sigma_j}+\frac{1}{\sigma_i\sigma_j}\int_{-\infty}^{+\infty}\int_{-\infty}^{+\infty}F^{-1}_{i}[\Phi(u_i)]F^{-1}_{j}[\Phi(\rho_zu_i+\sqrt{1-\rho_z^2}u_j)]\phi(u_i)\phi(u_j) du_idu_j.
\end{equation}
Employ a two-fold Gauss-Hermite quadrature with $m$ points to calculate the integral in Eq.\eqref{ppp}, and use the bisection method to determine $\rho_z$, then, the calculation times of $F^{-1}_{i}[\Phi(\cdot)]$ is $m$, and the calculation times of $F^{-1}_{j}[\Phi(\cdot)]$ would be $Tm^2$,  thus, the calculation times of $F^{-1}_{i}[\Phi(\cdot)]$ and $F^{-1}_{j}[\Phi(\cdot)]$ are ($Tm^2+m$).

The linear search method is developed under the assumption that the double integral can be accurately calculated by Gauss-Hermite quadrature, that's to say, the functions  $F^{-1}_{i}[\Phi(\cdot)]$ and $F^{-1}_{j}[\Phi(\cdot)]$ can be well approximated by Eq.\eqref{Polynomial22} ($n=m$). Then, an $m$th-order polynomial in Eq.\eqref{wwww} can also be used to approximate the function relationship between $\rho_x$ and $\rho_z$, and ($m+1$) values of $a_{i,k_i}$ and ($m+1$) values of $a_{j,k_j}$ ($k_i,k_j=0,\dots,m$) should be calculated by integrals in Eq.\eqref{vcc}  with respect to $F_i^{-1}[\Phi(\cdot)]$ and  $F_j^{-1}[\Phi(\cdot)]$ respectively.

For  an $m$th-order Hermite polynomial, all the  integrals  in Eq.\eqref{vcc} can be accurately calculated by a Gauss-Hermite quadrature with ($m+1$) points, which has an algebraic accuracy with degree ($2m+1$). For the proposed method, the calculation times of $F_i^{-1}[\Phi(\cdot)]$ and  $F_j^{-1}[\Phi(\cdot)]$ is ($m+1$) respectively, totaling ($2m+2$) times. Compared to the linear search method, ($Tm^2-m-2$) calculation times are saved. For many distributions, the calculation of  $F^{-1}[\Phi(\cdot)]$ involves numerical approaches, and the proposed method would be  more efficient than the linear search method.

\subsection{Discrete case}\label{CompareDiscrete}
If $x_i$ and $x_j$ are discrete variables, suppose the support of $x_i$ is $\{X_{i,k_i}\}$ ($k_i=1,\dots,N_i$), the support of $x_j$ is $\{X_{j,k_j}\}$ ($k_j=1,\dots,N_j$). By the marginal transformation in Eq.\eqref{CopulaD2}, $\{Z_{i,k_i}\}$ ($k_i=0,1,\dots,N_i$) and  $\{Z_{j,k_j}\}$ ($k_j=0,1,\dots,N_j$) are obtained, whereby Eq.\eqref{NatafEq345} is decomposed into  a sum    in Eq.\eqref{finall-2}. If $\rho_z$ is determined by a bisection method, it needs to calculate $\Phi(z_i,z_j,\rho_z)$ $4TN_iN_j$ times.

For the proposed method, if an $n$th-order Taylor series in Eq.\eqref{Taylor} is employed, it requires to evaluate the values of  $0-(n-1)$th order Hermite polynomials and $\phi(\cdot)$ at   $(N_i+N_j+2)$ points of $\{Z_{i,k_i}\}$ and  $\{Z_{j,k_j}\}$ respectively (see Eq.\eqref{DDDD02} and Eq.\eqref{dsdv2222s}), and the calculation times of   Hermite polynomials and $\phi(\cdot)$ are $2(n-1)(N_i+N_j+2)$ respectively (note that the 0th-order Hermite polynomials is 1).

Because the calculation  of Hermite polynomials and $\phi(\cdot)$ is   more efficient than the  calculation of $\Phi(z_i,z_j,\rho_z)$,   when $N_i$ or $N_j$ is large,  a lot of computational time can be saved by the proposed method.

 \subsection{Mixed case}
If a discrete variable and a continuous variable are involved, let $x_i$ be the discrete one. Suppose the support of $x_i$ is $\{X_{i,k_i}\}$ ($k_i=1,\dots,N_i$). According to Eq.\eqref{CopulaD2} and Eq.\eqref{ppp}, it has:
\begin{equation}\label{vcc33}
\begin{split}
\rho_x=&-\frac{\mu_1\mu_2}{\sigma_1\sigma_2}+\frac{1}{\sigma_1\sigma_2}\sum_{k_i=1}^{N_i}X_{i,k_i}\int_{-\infty}^{+\infty}\Bigg\{\int_{Z_{i,k_{i}-1}}^{Z_{i,k_{i}}}F^{-1}_{j}[\Phi(\rho_zu_i+\sqrt{1-\rho_z^2}u_j)]\phi(u_i)du_i\Bigg\}\\
&\phi(u_j) du_j
\end{split}
\end{equation}
Suppose the outer integral is calculated by an $m_1$-point Gauss-Hermite quadrature, the inner integral is calculated by an $m_2$-point Gauss-Legendre quadrature, the calculation times of $F_j^{-1}[\Phi(\cdot)]$ would be $TN_im_1m_2$ for bisection method.

If an $n$th-order Taylor series in Eq.\eqref{Taylor} is employed, it requires to calculate $F_j^{-1}[\Phi(\cdot)]$ ($n+1$) times,  $0-n$th order Hermite polynomials $n(N_i+1)$ times and $\phi(\cdot)$  $n(N_i+1)$ times (suppose the integral in Eq.\eqref{mix5} is calculated by Gauss-Hermite quadrature with ($n+1$) points).

\section{Examples}
Suppose $x_i$ and $x_j$ both follow Beta distribution $Beta(2,3)$. Several values of $\rho_x$ are selected, the corresponding values of $\rho_z$ are determined by  linear search method in\cite{Nataf4} with error bound $\varepsilon=10^{-3}$, interpolation method in\cite{NatafXiao} and proposed method respectively.  The integral in Eq.\eqref{ppp} is calculated by a two-fold Gauss-Hermite quadrature with 11 points. The Monte Carlo(MC) method  with $10^6$ points in\cite{NatafXiao} is employed to provide benchmark.  Along with computational time, the results are summarized in Table \ref{continuous}.

 \begin{table}[!htb]
\setlength{\abovecaptionskip}{0pt}
\centering
\caption{The values of $\rho_z$ for $Beta(2,3)\thicksim Beta(2,3)$}
\label{continuous}
{\begin{tabular}[l]{@{}ccccccccc}
\hline
 \multirow{2}*{$\rho_x$ }&\multicolumn{4}{c}{$Beta(2,3)\thicksim Beta(2,3)$}&\\
 \cline{2-5}
&Benchmark& Linear search& Interpolation& Proposed method\\
\hline
$-0.9$& $-0.914$& $-0.915$& $-0.914$& $-0.914$  \\
$-0.6$& $-0.611$& $-0.611$& $-0.611$& $-0.611$  \\
$-0.3$& $-0.306$& $-0.306$& $-0.306$& $-0.306$   \\
$0.3$&  $0.304$&  $0.304$&  $0.304$&  $0.304$   \\
$0.6$&  $0.606$&  $0.606$&  $0.606$&  $0.606$ \\
$0.9$& $0.904$& $0.903$& $0.903$& $0.903$\\
\hline
Time (s)&$-$&$12.2$ ($8795$)&$1.48$ ($1100$)&$0.015$ ($11$)\\
\hline
\end{tabular}}
\end{table}
The numerical experiment is performed  in MATLAB on a 2.3 GHz Intel Core i3-2350M computer with 3 GB of RAM. As discussed in Section \ref{CompareContinuous}, the efficiency of these three methods links directly to the calculation times of the function $F^{-1}[\Phi(\cdot)]$, which are presented in the brackets in  the last row of Table \ref{continuous}. All three methods yield results of the same level of accuracy, but the proposed method is   more efficient than other two methods.



Here, two example associated with the discrete case is performed. Suppose $x_i$ and $x_j$ both follow Binomial distribution $B(n,p)$. Two scenarios: $n=2,~p=0.2$ and $n=20,~p=0.2$ are considered. Using the method in Section \ref{DetermineDegree} ($\Delta n=2$, $\delta=10^{-4}$), for  the case of $B(2,0.2)$,   it takes $0.014$ seconds to determine that a $23$th-order Taylor series in Eq.\eqref{Taylor}  should be employed  to approximated $G(\cdot)$; for  the case of $B(20,0.2)$, a $3$th-order Taylor series should be employed, and the computational time is $0.068$ seconds.   Choose several values of $\rho_x$, $\rho_z$ are calculated by the proposed method and NI1 method in \cite{NatafDDK}. With benchmark from MC method ($10^6$ points), the results are presented in Table \ref{table11ddd}.

\begin{table}[!htb]
\setlength{\abovecaptionskip}{0pt}
\centering
\caption{The values of $\rho_z$ for $B(2,0.2) \thicksim B(2,0.2)$ and $B(20,0.2) \thicksim B(20,0.2)$}
\label{table11ddd}
{\begin{tabular}[l]{@{}cccccccc}
\hline
 \multirow{2}*{$\rho_x$ }&\multicolumn{3}{c}{$B(2,0.2) \thicksim B(2,0.2)$} &\multirow{2}*{$\rho_x$ }&\multicolumn{3}{c}{$B(20,0.2) \thicksim B(20,0.2)$}\\
 \cline{2-4}\cline{6-8}
&Benchmark&$n=23$&NI1&&Benchmark&$n=3$&NI1\\
\hline
$-0.5$&$-0.947$&$-0.946$&$-0.946$    &$-0.9$ &$-0.939$ &$-0.938$ &$-0.938$\\
$-0.3$&$-0.501$&$-0.501$&$-0.501$    &$-0.6$&$-0.624$&$-0.624$&$-0.624$\\
$-0.2$&$-0.322$&$-0.322$&$-0.322$    &$-0.3$&$-0.311$&$-0.311$&$-0.311$\\
$0.3$&$0.418$&$0.418$&$0.418$    &$0.3$&$0.310$&$0.310$&$0.310$\\
$0.6$&$0.769$&$0.769$&$0.769$    &$0.6$&$0.618$&$0.618$&$0.618$\\
$0.8$&$0.944$&$0.943$&$0.943$    &$0.9$&$0.925$&$0.925$&$0.925$\\
\hline
Time (s)&$-$&$0.014$&$0.21$&Time (s)&$-$&$ 0.068$&$10.5$\\
\hline
\end{tabular}}
\end{table}

 For the case of $B(2,0.2)$, both methods are efficient, but as discussed in Section \ref{CompareDiscrete}, the computational time of NI1 method increases sharply for the case of $B(20,0.2)$.

Finally, two examples for the mixed case are performed. Suppose $x_i$ follows Binomial distribution $B(2,0.2)$ or  $B(20,0.2)$, $x_j$ follows Beta distribution $Beta(2,3)$.   MC method with $10^6$ points are employed to provide benchmark.  Except for the proposed method, a bisection search method based on Eq.\eqref{vcc33} is also employed to determine $\rho_z$, and the inner integral is calculated by an $11$-point  Gauss-Legendre quadrature, the outer integral is calculated by  an $11$-point  Gauss-Hermite quadrature.  The error bound is $\varepsilon=10^{-3}$.  The results are presented in Table \ref{table1xxx1ddd}.

\begin{table}[!htb]
\setlength{\abovecaptionskip}{0pt}
\centering
\caption{The values of $\rho_z$ for $B(2,0.2) \thicksim Beta(2,3)$ and $B(20,0.2) \thicksim Beta(2,3)$}
\label{table1xxx1ddd}
{\begin{tabular}[l]{@{}cccccccc}
\hline
 \multirow{2}*{$\rho_x$ }&\multicolumn{3}{c}{$B(2,0.2) \thicksim Beta(2,3)$} &\multirow{2}*{$\rho_x$ }&\multicolumn{3}{c}{$B(20,0.2) \thicksim Beta(2,3)$}\\
 \cline{2-4}\cline{6-8}
&Benchmark&$n=7$&Eq.\eqref{vcc33}  &&Benchmark&$n=5$&Eq.\eqref{vcc33} \\
\hline
$-0.7$&$-0.890$&$-0.889$&$-0.889$    &$-0.9$ &$-0.928$ &$-0.929$&$-0.929$\\
$-0.5$&$-0.632$&$-0.632$&$-0.631$    &$-0.6$&$-0.618$&$-0.618$&$-0.618$\\
$-0.3$&$-0.377$&$-0.377$&$-0.376$    &$-0.3$&$-0.309$&$-0.309$&$-0.309$\\
$0.3$&$0.366$&$0.366$&$0.367$   &$0.3$&$0.308$&$0.308$&$0.307$\\
$0.5$&$0.603$&$0.603$&$0.603$    &$0.6$&$0.613$&$0.613$&$0.613$\\
$0.8$&$0.945$&$0.945$&$0.944$    &$0.9$&$0.916$&$0.916$&$0.916$\\
\hline
Time (s)&$-$&$0.023$&$32.9$&Time (s)&$-$&$ 0.053$&$222.5$\\
\hline
\end{tabular}}
\end{table}
Compared to the former two examples, the linear search method takes a lot more time, because the calculation of $F^{-1}_j[\Phi(\cdot)]$   has been performed $21~054$ times for the case of $B(2,0.2) \thicksim Beta(2,3)$ and  $147~378$ times for the case of $B(20,0.2) \thicksim Beta(2,3)$.

\section{Conclusion}
This paper attempts to  determine the equivalent correlation coefficient $\rho_z$ for Gaussian copula. For  the continuous random variable, the marginal transformation is approximated by a weighted sum of Hermite polynomials; for the discrete random variable, the marginal transformation is decomposed into  piecewise continuous ones. Using Mehler's formula and Taylor series, a polynomial of $\rho_z$ is developed to approximate the function relationship between $\rho_z$ and  $\rho_x$.  The numerical examples show the efficiency and accuracy of the proposed method.

\section{Appendix}
Using Hermite polynomials and Mehler's formula, the function relationship between $\rho_x$ and $\rho_z$ can be determined analytically for a few cases (see Table \ref{gt}).
\begin{table}[!h]\tiny
\setlength{\abovecaptionskip}{0pt}
\renewcommand{\arraystretch}{1.5}
\centering
\caption{The function relationship between $\rho_z$ and $\rho_x$}
\label{gt}
{\begin{tabular}[l]{@{}lccl}
\hline
&$x_1$&$x_2$&\\
\hline
\nameref{CASE1}& Uniform distribution $U(0, 1)$& Uniform distribution $U(0, 1)$&$\rho_z=2sin\left(\frac{\pi}{6}\rho_x\right)$\\
\nameref{CASE2} &Uniform distribution $U(0, 1)$& Binomial distribution $B(1,0.5)$&$\rho_z=\sqrt{2}sin\left(\frac{\pi}{2\sqrt{3}}\rho_x\right)$\\
\nameref{CASE3}& Uniform distribution $U(0, 1)$& Normal distribution $N(0, 1)$&$\rho_z=\sqrt{\frac{\pi}{3}}\rho_x$\\
\nameref{CASE4}& Uniform distribution $U(0, 1)$& Lognormal distribution $lnN(\mu_2,\sigma_2^2)$&$\rho_x=\frac{2\sqrt{3}\Phi(\frac{\sigma_2\rho_z}{\sqrt{2}})-\sqrt{3}}{\sqrt{e^{\sigma_2^2}-1}}$\\
\nameref{CASE5}& Binomial distribution $B(1,0.5)$& Binomial distribution $B(1,0.5)$&$\rho_z=sin\left(\frac{\pi}{2}\rho_x\right)$\\
\nameref{CASE6}& Binomial distribution $B(1,0.5)$& Normal distribution $N(0,1)$&$\rho_z=\sqrt{\frac{\pi}{2}}\rho_x$\\
\nameref{CASE7}& Binomial distribution $B(1,0.5)$& Lognormal distribution $lnN(\mu_2,\sigma_2^2)$&$\rho_x=\frac{2\Phi(\sigma_2\rho_z)-1}{\sqrt{e^{\sigma_2^2}-1}}$\\
\nameref{CASE8}& Normal distribution $N(0,1)$& Lognormal distribution $lnN(\mu_2,\sigma_2^2)$&$\rho_x=\frac{\sigma_2 \rho_z}{\sqrt{e^{\sigma_2^2}-1}}$\\
\nameref{CASE9}& Lognormal distribution $lnN(\mu_1,\sigma_1^2)$& Lognormal distribution $lnN(\mu_2,\sigma_2^2)$&$\rho_x=\frac{e^{\sigma_1\sigma_2\rho_z}-1}{\sqrt{(e^{\sigma_1^2}-1)(e^{\sigma_2^2}-1)}}$\\
\hline
\end{tabular}}
\end{table}

\subsection{Hermite polynomials expansion of some functions}

For Hermite polynomials, the following equations hold:




\begin{equation}\label{3}
\begin{split}
\int_{-\infty}^{+\infty}H_{k}(z)\Phi(z)\phi(z)dz=\Bigg\{
             \begin{array}{lll}
                            \frac{1}{2} ~~~~~~~~~~ &k=0\\
                            0~~~~~~~~~&k=2n+2\\
              \frac{(-1)^n(2n)!}{\sqrt{4\pi}4^nn!}  &k=2n+1.\\
             \end{array}
             \end{split}
\end{equation}

\begin{equation} \label{1}
\begin{split}
\int_{0}^{\infty}H_{k}(z)\phi(z)dz=\Bigg\{
             \begin{array}{lll}
                            \frac{1}{2} ~~~~~~~~~~~~~~~~~~ &k=0\\
                            0~~~~~~~~~~~~~~~~~~&k=2n+2\\
           \frac{(-1)^n(2n)!}{\sqrt{2\pi}2^nn!} ~~~~~~~ &k=2n+1.\\
             \end{array}\\
             \end{split}
\end{equation}

\begin{equation}\label{1e}
\int_{-\infty}^{+\infty}e^{az}H_{k}(z)\phi(z)dz=e^{\frac{a^2}{2}}a^k.~~~~~~~~~~~~~~~~~~~~~~~~~
\end{equation}

Using Eq.\eqref{3},  an Hermite polynomial expansion of $\Phi(z)$ can be obtained:
\begin{equation}\label{a1}
\Phi(z)=\sum_{k=0}^{\infty}\frac{\int_{-\infty}^{\infty}H_k(z)\Phi(z)\phi(z)dz}{k!}H_{k}(z)=\frac{1}{2}+\sum_{n=0}^{\infty}\frac{(-1)^nH_{2n+1}(z)}{\sqrt{4\pi}(2n+1)4^nn!}.
\end{equation}

Using Eq.\eqref{1e}, the Hermite polynomial expansion of $e^{az}$ is:
\begin{equation}\label{c4}
e^{az}=\sum_{k=0}^{\infty}\frac{ \int_{-\infty}^{+\infty}e^{az}H_k(z)\phi(z)dz }{k!}H_k(z)=e^{\frac{a^2}{2}}\cdot\sum_{k=0}^{\infty}\frac{ a^k}{k!}H_k(z).
\end{equation}
\subsubsection{Proof of  Eq.\eqref{3}}
 Consider the Taylor series of $\phi(z)$:
\begin{equation}
\begin{split}
\phi(z)=\frac{1}{\sqrt{2\pi}}e^{-\frac{z^2}{2}}=\frac{1}{\sqrt{2\pi}}\sum_{n=0}^{\infty}\frac{(-1)^nz^{2n}}{n!2^n},
\end{split}
\end{equation}
then,
\begin{equation}\label{fff}
\begin{split}
\Phi(z)=\int_{-\infty}^z\phi(t)dt=\int_{-\infty}^0\phi(t)dt+\int_{0}^z\phi(t)dt=\frac{1}{2}+\frac{1}{\sqrt{2\pi}}\sum_{n=0}^{\infty}\frac{(-1)^nz^{2n+1}}{(2n+1)2^nn!}.
\end{split}
\end{equation}
The generating function of Hermite polynomials is:
\begin{equation}\label{Generating}
e^{tz-\frac{t^2}{2}}=\sum_{k=0}^{\infty}H_k(z)\frac{t^k}{k!}.
\end{equation}
Then:
\begin{equation} \label{cc2}
\int_{-\infty}^{+\infty} e^{tz-\frac{t^2}{2}}\Phi(z)\phi(z)dz=\int_{-\infty}^{+\infty} \sum_{k=0}^{\infty}H_k(z)\frac{t^k}{k!}\cdot \Phi(z)\phi(z)dz=\sum_{k=0}^{\infty}\int_{-\infty}^{+\infty} H_k(z)\Phi(z)\phi(z)dz\cdot \frac{t^k}{k!}.
\end{equation}
On the other hand:
\begin{equation} \label{vvvv2}
\begin{split}
\int_{-\infty}^{+\infty}  e^{tz-\frac{t^2}{2}}\Phi(z)\phi(z)dz=&\int_{-\infty}^{+\infty} \Phi(z)\phi(z-t)dz= \int_{-\infty}^{+\infty} \Phi(u+t)\phi(u)du= \Phi\left(\frac{t}{\sqrt{2}}\right).
\end{split}
\end{equation}
The last step is due to the formula 10010.8 in \cite{owen1980table}.
Then, using Eq.\eqref{fff}, Eq.\eqref{vvvv2} can be expressed as:
\begin{equation} \label{vvvv4}
\begin{split}
\int_{-\infty}^{+\infty}  e^{tz-\frac{t^2}{2}}\Phi(z)\phi(z)dz=\Phi\left(\frac{t}{\sqrt{2}}\right)=
\frac{1}{2}+\frac{1}{\sqrt{4\pi}}\sum_{n=0}^{\infty}\frac{(-1)^nt^{2n+1}}{(2n+1)4^nn!},
\end{split}
\end{equation}
According to Eq.\eqref{cc2} and Eq.\eqref{vvvv4}, it has:
\begin{equation} \label{vvvv5}
\begin{split}
\sum_{k=0}^{\infty}\int_{-\infty}^{+\infty} H_k(z)\Phi(z)\phi(z)dz\cdot \frac{t^k}{k!}=
\frac{1}{2}+\frac{1}{\sqrt{4\pi}}\sum_{n=0}^{\infty}\frac{(-1)^nt^{2n+1}}{(2n+1)4^nn!}.
\end{split}
\end{equation}
Match the coefficient of $t^k$ ($k=0, 2n, 2n+1$), Eq.\eqref{3} can be obtained.

\subsubsection{Proof of  Eq.\eqref{1} }
The proof of Eq.\eqref{1} is similar:
\begin{equation} \label{cc}
\int_{0}^{\infty} e^{tz-\frac{t^2}{2}}\phi(z)dz=\int_{0}^{\infty}\sum_{k=0}^{\infty}H_k(z)\frac{t^k}{k!}\phi(z)dz=\sum_{k=0}^{\infty}\int_{0}^{\infty}H_k(z)\phi(z)dz\cdot \frac{t^k}{k!}.
\end{equation}
The left hand side of Eq.\eqref{cc} can be calculated as:
\begin{equation} \label{vvvv}
\begin{split}
\int_{0}^{\infty} e^{tz-\frac{t^2}{2}}\phi(z)dz=\int_{0}^{\infty}\phi(z-t)dz= \int_{-t}^{\infty}\phi(u)du=1-\Phi(-t).
\end{split}
\end{equation}
Substitute Eq.\eqref{fff} and Eq\eqref{cc}  into Eq.\eqref{vvvv}:
\begin{equation}
\sum_{k=0}^{\infty}\int_{0}^{\infty}H_k(z)\phi(z)dz\cdot \frac{t^k}{k!}=
\frac{1}{2}+\frac{1}{\sqrt{2\pi}}\sum_{n=0}^{\infty}\frac{(-1)^nt^{2n+1}}{(2n+1)2^nn!}.
\end{equation}
Match the coefficient of $t^k$, Eq.\eqref{1} can be obtained.
\subsubsection{Proof of   Eq.\eqref{1e}}
Using the generating function  in Eq.\eqref{Generating}, Eq.\eqref{1e} can also be proved.
\begin{equation}
 \int_{-\infty}^{+\infty}e^{az}e^{tz-\frac{t^2}{2}}\phi(z) dz= \sum_{k=0}^{\infty}\int_{-\infty}^{+\infty}e^{az}H_k(z)\phi(z)dz\frac{t^k}{k!},
\end{equation}
and
\begin{equation}
 \int_{-\infty}^{+\infty}e^{az}e^{tz-\frac{t^2}{2}}\phi(z) dz=  \int_{-\infty}^{+\infty} e^{az}\phi(z-t)dz=e^{at} \int_{-\infty}^{+\infty} e^{au}\phi(u)du=e^{at}\cdot e^{\frac{a^2}{2}}.
\end{equation}
The last step is due to the formula 1000n0 in \cite{owen1980table}. Then:
\begin{equation}
\sum_{k=0}^{\infty}\int_{-\infty}^{+\infty}e^{az}H_k(z)\phi(z)dz\frac{t^k}{k!}=e^{at}\cdot e^{\frac{a^2}{2}}=e^{\frac{a^2}{2}}\sum_{k=0}^{\infty} \frac{a^k t^k}{k!}.
\end{equation}
Match the coefficient of $t^k$, Eq.\eqref{1e} can be obtained.

Although it may be a little out the scope of this paper,  following this idea, the Hermite polynomial expansions of several elementary functions are obtained (Table \ref{ww1ww}). A point worth noting is that all these functions are closely related to the exponential function, and   Hermite polynomial expansions of these functions are similar to their Taylor expansions.

\begin{table}[!h]
\setlength{\abovecaptionskip}{0pt}
\renewcommand{\arraystretch}{1.3}
\centering
\caption{The Hermite polynomial expansions of several functions}
\label{ww1ww}
{\begin{tabular}[l]{@{}lll}
\hline

\hline
Functions&Hermite polynomial expansions&Taylor expansions\\
\hline
$\Phi(ax)$&$\frac{1}{2}+\frac{1}{\sqrt{2\pi}}\sum\limits_{n=0}^{\infty}\left(\frac{a}{\sqrt{1+a^2}}\right)^{2n+1}\cdot \frac{(-1)^nH_{2n+1}(x)}{(2n+1)2^nn!}$&$\frac{1}{2}+\frac{1}{\sqrt{2\pi}}\sum\limits_{n=0}^{\infty}\frac{(-1)^na^{2n+1}x^{2n+1}}{(2n+1)2^nn!}$\\
$\phi(ax)$&$\frac{1}{\sqrt{2\pi}a}\sum\limits_{n=0}^{\infty}\left(\frac{a}{\sqrt{1+a^2}}\right)^{2n}\cdot \frac{(-1)^nH_{2n}(x)}{2^nn!}$& $\frac{1}{\sqrt{2\pi}}\sum\limits_{n=0}^{\infty}\frac{(-1)^na^{2n}x^{2n}}{2^nn!}$\\
$e^{ax}$&$e^{\frac{a^2}{2}}\sum\limits_{n=0}^{\infty}\frac{a^nH_{n}(x)}{n!}$&$\sum\limits_{n=0}^{\infty}\frac{a^{n}x^{n}}{n!}$\\
$sinh(ax)$&$e^{\frac{a^2}{2}}\sum\limits_{n=0}^{\infty}\frac{a^{2n+1}H_{2n+1}(x)}{(2n+1)!}$&$\sum\limits_{n=0}^{\infty}\frac{a^{2n+1}x^{2n+1}}{(2n+1)!}$\\
$cosh(ax)$&$e^{\frac{a^2}{2}}\sum\limits_{n=0}^{\infty}\frac{a^{2n}H_{2n}(x)}{(2n)!}$&$\sum\limits_{n=0}^{\infty}\frac{a^{2n}x^{2n}}{(2n)!}$\\
$sin(ax)$&$e^{-\frac{a^2}{2}}\sum\limits_{n=0}^{\infty}\frac{(-1)^na^{2n+1}H_{2n+1}(x)}{(2n+1)!}$&$\sum\limits_{n=0}^{\infty}\frac{(-1)^na^{2n+1}x^{2n+1}}{(2n+1)!}$\\
$cos(ax)$&$e^{-\frac{a^2}{2}}\sum\limits_{n=0}^{\infty}\frac{(-1)^na^{2n}H_{2n}(x)}{(2n)!}$&$\sum\limits_{n=0}^{\infty}\frac{(-1)^na^{2n}x^{2n}}{(2n)!}$\\

\hline
\end{tabular}}
\end{table}

With results in Table \ref{ww1ww}, some interesting formulae can be derived:
\begin{equation}
\begin{split}
&\int_{-\infty}^{\infty}cos(az)\phi(z)dz= e^{-\frac{a^2}{2}}\\
&\int_{-\infty}^{\infty}\int_{-\infty}^{\infty}sin(a_1z_1)sin(a_2z_2)\phi(z_1,z_2,\rho_z)dz_1dz_2=sinh(a_1a_2\rho_z)\cdot e^{-\frac{a_1^2+a_2^2}{2}}\\
&\int_{-\infty}^{\infty}\int_{-\infty}^{\infty}cos(a_1z_1)cos(a_2z_2)\phi(z_1,z_2,\rho_z)dz_1dz_2=cosh(a_1a_2\rho_z)\cdot e^{-\frac{a_1^2+a_2^2}{2}}\\
&\int_{-\infty}^{\infty}\int_{-\infty}^{\infty}cos(a_1z_1-a_2z_2)\phi(z_1,z_2,\rho_z)dz_1dz_2= e^{-\frac{a_1^2-2\rho_za_1a_2+a_2^2}{2}}\\
&\int_{-\infty}^{\infty}\int_{-\infty}^{\infty}cos(a_1z_1+a_2z_2)\phi(z_1,z_2,\rho_z)dz_1dz_2= e^{-\frac{a_1^2+2\rho_za_1a_2+a_2^2}{2}}\\
\end{split}
\end{equation}
Using these formulae, the Fourier series for $\phi(z)$ on interval $[-T,T]$ can be obtained analytically. If $T$ is sufficiently large, then:
\begin{equation}
\int_{-T}^{T}cos(az)\phi(z)dz\simeq\int_{-\infty}^{\infty}cos(az)\phi(z)dz= e^{-\frac{a^2}{2}},\\
\end{equation}
and the error between these two integrals can be bounded:
\begin{equation}
\begin{split}
e^{-\frac{a^2}{2}}-\int_{-T}^{T}cos(az)\phi(z)dz=&2\int_{T}^{\infty}cos(az)\phi(z)dz<2\int_{T}^{\infty}\phi(z)dz=2[1-\Phi(T)].
\end{split}
\end{equation}
Following the routine of calculating  coefficients of Fourier series, the Fourier series for $\phi(z)$ on $[-T\leq z \leq T]$ is:
\begin{equation}
\phi(z)\simeq \frac{1}{2T}+\frac{1}{T}\sum_{k=1}^{n}e^{-\frac{k^2\pi^2}{2T^2}}cos\left(\frac{k\pi}{T}z\right).\\
\end{equation}
The extension to $n$-dimensional standard normal distribution is straightforward, the Fourier series for $\phi(z_1,z_2,\rho_z)$ on $[-T\leq z_1,z_2 \leq T]$  is:
\begin{equation}
  \begin{split}
\phi(z_1,z_2,\rho_z)\simeq& \frac{1}{4T^2}+\frac{1}{2T^2}\sum_{k_1=1}^{n}e^{-\frac{k_1^2\pi^2}{2T^2}}cos\left(\frac{k_1\pi}{T}z_1\right)+\frac{1}{2T^2}\sum_{k_2=1}^{n}e^{-\frac{k_2^2\pi^2}{2T^2}}cos\left(\frac{k_2\pi}{T}z_2\right)+ \\ &\frac{1}{T^2}\sum_{k_1=1}^n\sum_{k_2=1}^n e^{-\frac{k_1^2+k_2^2}{2T^2}\pi^2}\Big[cosh\left(\frac{k_1k_2}{T^2}\pi^2\rho_z\right)cos(k_1z_1)cos(k_2z_2)+\\
&sinh\left(\frac{k_1k_2}{T^2}\pi^2\rho_z\right)sin(k_1z_1)sin(k_2z_2)\Big]\\
\end{split}
\end{equation}

\subsection{Proof of the formulae in Table \ref{gt}}
The mean and standard deviation of  probability distributions in Table \ref{gt} are presented here (Table \ref{gt2}).
\begin{table}[!h]
\setlength{\abovecaptionskip}{0pt}
\renewcommand{\arraystretch}{1.3}
\centering
\caption{The mean and standard deviation of listed distributions}
\label{gt2}
{\begin{tabular}[l]{@{}cccl}
\hline
&Mean&Standard deviation\\
\hline
Uniform distribution $U(0, 1)$&$\frac{1}{2}$&$\frac{1}{2\sqrt{3}}$\\
Binomial distribution $B(1,0.5)$&$\frac{1}{2}$&$\frac{1}{2}$\\
Normal distribution $N(0,1)$&$0$&$1$\\
Lognormal distribution $lnN(\mu,\sigma^2)$&$e^{\mu+\frac{\sigma^2}{2}}$&$\sqrt{e^{\sigma^2}-1}e^{\mu+\frac{\sigma^2}{2}}$\\
\hline
\end{tabular}}
\end{table}

For Case \uppercase\expandafter{\romannumeral1}, Case \uppercase\expandafter{\romannumeral2} and Case \uppercase\expandafter{\romannumeral5}, the Taylor series of $arcsin(x)$ is essential:
\begin{equation}
arcsin(x)=\sum_{n=0}^{\infty}\frac{(2n)!x^{2n+1}}{4^n(n!)^2(2n+1)}.
\end{equation}


\subsubsection{Case \uppercase\expandafter{\romannumeral1}}\label{CASE1}
For the case of $U(0,1)$ and $U(0,1)$, using Eq.\eqref{a1}, it has:
\begin{equation}\label{27}
\begin{split}
\frac{1}{12}\rho_x+\frac{1}{4}=&E[\Phi(z_1)\Phi(z_2)]\\
=&E\left[\left(\frac{1}{2}+\sum_{n_i=0}^{\infty}\frac{(-1)^{n_i}H_{2n_i+1}(z_1)}{\sqrt{4\pi}(2n_i+1)4^{n_i}n_i!} \right)\left(\frac{1}{2}+\sum_{n_j=0}^{\infty}\frac{(-1)^{n_j}H_{2n_j+1}(z_2)}{\sqrt{4\pi}(2n_j+1)4^{n_j}n_j!} \right)\right]\\
=&\frac{1}{4}+E\left[\left(\sum_{n_i=0}^{\infty}\frac{(-1)^{n_i}H_{2n_i+1}(z_1)}{\sqrt{4\pi}(2n_i+1)4^{n_i}n_i!} \right)\left(\sum_{n_j=0}^{\infty}\frac{(-1)^{n_j}H_{2n_j+1}(z_2)}{\sqrt{4\pi}(2n_j+1)4^{n_j}n_j!} \right)\right]\\
=&\frac{1}{4}+\frac{1}{{4\pi}}\sum_{n_i=0}^{\infty}\sum_{n_j=0}^{\infty}E\left[\left(\frac{(-1)^{n_i}H_{2n_i+1}(z_1)}{(2n_i+1)4^{n_i}n_i!} \right)\left(\frac{(-1)^{n_j}H_{2n_j+1}(z_2)}{(2n_j+1)4^{n_j}n_j!} \right)\right]\\
\end{split}
\end{equation}
The third step is due to   that, if $z$ is a standard normal variable, $E[H_{2n+1}(z)]=0$ (see Eq.\eqref{AnaH}):
\begin{equation}\label{AnaH}
\begin{split}
H_{k}(z)=\Bigg\{
             \begin{array}{ll}
             (2n)!\sum\limits_{s=0}^n\frac{(-1)^{n-s}}{(n-s)!2^{n-s}}\cdot\frac{z^{2s}}{(2s)!}~~~ ~~~~~~~~~~ k=2n\\
              (2n+1)!\sum\limits_{s=0}^n\frac{(-1)^{n-s}}{(n-s)!2^{n-s}}\cdot\frac{z^{2s+1}}{(2s+1)!}~~~~~ k=2n+1.\\
             \end{array}\\
             \end{split}
\end{equation}

 With Mehler's formula in Eq.\eqref{doublenormal2}, Eq.\eqref{27} becomes:
\begin{equation}
\begin{split}
\rho_x=&\frac{3}{{\pi}}\sum_{n_i=0}^{\infty}\sum_{n_j=0}^{\infty}E\left[\left(\frac{(-1)^{n_i}H_{2n_i+1}(z)}{(2n_i+1)4^{n_i}n_i!} \right)\left(\frac{(-1)^{n_j}H_{2n_j+1}(z)}{(2n_j+1)4^{n_j}n_j!} \right)\right]\\
=&\frac{3}{{\pi}}\sum_{n_i=0}^{\infty}\sum_{n_j=0}^{\infty}\sum_{k=0}^{\infty}\frac{\rho_z^k}{k!}\int_{-\infty}^{+\infty}\frac{(-1)^{n_i}H_{2n_i+1}(z)}{(2n_i+1)4^{n_i}n_i!} \phi(z_1) H_k(z_1)dz_1\\
&\int_{-\infty}^{+\infty}\frac{(-1)^{n_j}H_{2n_j+1}(z)}{(2n_j+1)4^{n_j}n_j!} \phi(z_2) H_k(z_2)dz_2\\
\end{split}
\end{equation}
According to  Eq.\eqref{Hermite}, the integrals are not 0  if and only if $2n_i+1=k=2n_j+1$. Denote $k=2n+1$, then:
\begin{equation}
\begin{split}
\rho_x
=&\frac{3}{{\pi}}\sum_{n=0}^{\infty}\frac{\rho_z^{2n+1}}{(2n+1)!}\cdot\frac{(-1)^{n}(2n+1)!}{(2n+1)4^{n}n!} \cdot \frac{(-1)^{n}(2n+1)!}{(2n+1)4^{n}n!}=\frac{3}{{\pi}}\sum_{n=0}^{\infty}\frac{\rho_z^{2n+1}(2n)!}{2^{2n}4^{n}(n!)^2(2n+1)} \\
=&\frac{6}{{\pi}}\sum_{n=0}^{\infty}\frac{(\frac{\rho_z}{2})^{2n+1}(2n)!}{4^{n}(n!)^2(2n+1)} =\frac{6}{{\pi}}arcsin(\frac{\rho_z}{2}),
\end{split}
\end{equation}
\begin{equation}\label{case1}
\rho_z=2sin\left(\frac{\pi}{6}\rho_x\right).
\end{equation}
Another two  proofs of Eq.\eqref{case1} can be found in \cite{case12,case11}.

\subsubsection{Case \uppercase\expandafter{\romannumeral2}}\label{CASE2}
For the case of $U(0,1)$ and $B(1,0.5)$, using Mehler's formula in Eq.\eqref{doublenormal2}, it has:
\begin{equation}
\begin{split}
\rho_x\frac{1}{4\sqrt{3}}+\frac{1}{4}=&\int_{-\infty}^{+\infty}\int_{0}^{+\infty} \Phi(z_1)\phi(z_1,z_2,\rho_z)dz_2dz_1\\
=&\sum_{k=0}^{\infty}\frac{\rho_z^k}{k!} \int_{-\infty}^{+\infty} \Phi(z_1) \phi(z_1) H_k(z_1) dz_1\cdot\int_{0}^{+\infty}\phi(z_2) H_k(z_2)dz_2.\\
\end{split}
\end{equation}
Using Eqs.\eqref{3}\eqref{1}, it has:
\begin{equation}
\begin{split}
\rho_x\frac{1}{4\sqrt{3}}+\frac{1}{4} =\frac{1}{4}+\sum_{n=0}^{\infty}\frac{(2n)!\rho_z^{2n+1}}{2\sqrt{2}2^n\pi4^n(n!)^2(2n+1)},
\end{split}
\end{equation}
and
\begin{equation}
\begin{split}
\rho_x=\frac{2\sqrt{3}}{\pi}\sum_{n=0}^{\infty}\frac{(2n)!\left(\frac{\rho_z}{\sqrt{2}}\right)^{2n+1}}{4^n(n!)^2(2n+1)}=\frac{2\sqrt{3}}{\pi}arcsin\left(\frac{\rho_z}{\sqrt{2}}\right),
\end{split}
\end{equation}
\begin{equation}\label{case22}
\rho_z=\sqrt{2}sin\left(\frac{\pi}{2\sqrt{3}}\rho_x\right).
\end{equation}

\subsubsection{Case \uppercase\expandafter{\romannumeral3}}\label{CASE3}
For the case of $U(0,1)$ and $N(0,1)$, using Eq.\eqref{Hermite}, Eq.\eqref{doublenormal2} and  Eq.\eqref{3}, it has:
\begin{equation}
\begin{split}
\rho_x\frac{1}{2\sqrt{3}}=&\int_{-\infty}^{+\infty}\int_{-\infty}^{+\infty} \Phi(z_1)z_2\phi(z_1,z_2,\rho_z)dz_2dz_1\\
=&\int_{-\infty}^{+\infty}\int_{-\infty}^{+\infty} \Phi(z_1)z_2\phi(z_1)\phi(z_2) \sum_{k=0}^{\infty}\frac{\rho_z^k}{k!}H_k(z_1)H_k(z_2)dz_2dz_1\\
=&\sum_{k=0}^{\infty}\frac{\rho_z^k}{k!}\cdot\int_{-\infty}^{+\infty}\Phi(z_1) H_k(z_1)\phi(z_1)dz_1 \int_{-\infty}^{+\infty} z_2  H_k(z_2)\phi(z_2) dz_2\\
=&\frac{\rho_z}{\sqrt{4\pi}}.\\
\end{split}
\end{equation}
Then:
\begin{equation}
\begin{split}
\rho_x=\sqrt{\frac{3}{\pi}}\rho_z \leftrightarrow \rho_z=\sqrt{\frac{\pi}{3}}\rho_x.
\end{split}
\end{equation}

\subsubsection{Case \uppercase\expandafter{\romannumeral4}}\label{CASE4}
For the case of $U(0,1)$ and $lnN(\mu_2,\sigma_2^2)$, using  Eq.\eqref{doublenormal2},  Eq.\eqref{3} and   Eq.\eqref{c4}, it has:
\begin{equation}
\begin{split}
&\rho_x\frac{ \sqrt{e^{\sigma_2^2}-1}e^{\mu_2+\frac{\sigma_2^2}{2}}}{2\sqrt{3}}+\frac{e^{\mu_2+\frac{\sigma_2^2}{2}} }{2} =\int_{-\infty}^{+\infty}\int_{-\infty}^{+\infty}
\Phi(z_1)e^{\mu_2+\sigma_2z_2}\phi(z_1,z_2,\rho_z)dz_1dz_2\\
=&e^{\mu_2+\frac{\sigma_2^2}{2}}\sum_{k_2=0}^{\infty}\sum_{k=0}^{\infty}\frac{\rho_z^k}{k!}\int_{-\infty}^{+\infty}
\Phi(z_1)H_{k}(z_1)\phi(z_1)dz_1\cdot \int_{-\infty}^{+\infty}
\frac{\sigma_2^{k_2}}{k_2!}H_{k_2}(z_2)H_{k}(z_2)\phi(z_2)dz_2\\
=&e^{\mu_2+\frac{\sigma_2^2}{2}}\sum_{k_2=0}^{\infty}\sum_{k=0}^{\infty}\frac{\rho_z^k}{k!}\left(\sum_{k=0}^{}\frac{1}{2}+ \sum_{k=2n+1}^{}\frac{(-1)^n(2n)!}{\sqrt{4\pi}4^nn!} \right)\cdot \int_{-\infty}^{+\infty}
\frac{\sigma_2^{k_2}}{k_2!}H_{k_2}(z_2)H_{k}(z_2)\phi(z_2)dz_2\\
=&e^{\mu_2+\frac{\sigma_2^2}{2}}\left(\frac{1}{2}+\sum_{n=0}^{\infty}\frac{(-1)^n(2n)!}{\sqrt{4\pi}4^nn!}\cdot\frac{\rho_z^{2n+1}}{(2n+1)!}\cdot\sigma_2^{2n+1} \right)\\
=&e^{\mu_2+\frac{\sigma_2^2}{2}}\Phi(\frac{\sigma_2\rho_z}{\sqrt{2}})\\
\end{split}
\end{equation}
The last step is due to Eq.\eqref{vvvv4}. Then:
\begin{equation}
\begin{split}
\rho_x=\frac{2\sqrt{3}\Phi(\frac{\sigma_2\rho_z}{\sqrt{2}})-\sqrt{3}}{\sqrt{e^{\sigma_2^2}-1}}.
\end{split}
\end{equation}

\subsubsection{Case \uppercase\expandafter{\romannumeral5}}\label{CASE5}
For the case of $B(1,0.5)$ and $B(1,0.5)$, using   Eq.\eqref{doublenormal2} and Eq.\eqref{1}, it has:
\begin{equation}
\begin{split}
\rho_x\frac{1}{4}+\frac{1}{4}=&\int_{0}^{+\infty}\int_{0}^{+\infty} \phi(z_1,z_2,\rho_z)dz_2dz_1\\
=&\sum_{k=0}^{\infty}\frac{\rho_z^k}{k!} \int_{0}^{+\infty}   \phi(z_1) H_k(z_1) dz_1\cdot\int_{0}^{+\infty}\phi(z_2) H_k(z_2)dz_2\\
=&\frac{1}{4}+\sum_{n=0}^{\infty} \frac{\rho_z^{2n+1}}{(2n+1)!}\cdot \frac{(-1)^n(2n)!}{\sqrt{2\pi}2^nn!}\cdot \frac{(-1)^n(2n)!}{\sqrt{2\pi}2^nn!},
\end{split}
\end{equation}
then:
\begin{equation}
\begin{split}
\rho_x=\frac{2}{\pi}\sum_{n=0}^{\infty}\frac{(2n)!\rho_z^{2n+1}}{4^n(n!)^2(2n+1)}=\frac{2}{\pi}arcsin(\rho_z),
\end{split}
\end{equation}
\begin{equation}\label{sin2}
\rho_z=sin\left(\frac{\pi}{2}\rho_x\right).
\end{equation}

Eq.\eqref{sin2} can also be proved in another way.  Consider the derivative of  ${\rho_x}$ with respect to ${\rho_z}$:
\begin{equation}\label{DDDD2}
\begin{split}
\frac{d\rho_x}{d\rho_z} =\frac{2}{\pi}\cdot \frac{1}{\sqrt{1-\rho_z^2}}\rightarrow \frac{d\rho_x}{d\rho_z} =\frac{2}{\pi}\cdot\frac{ d( arcsin(\rho_z))}{d\rho_z} .
\end{split}
\end{equation}
Then, $\rho_x=\frac{2}{\pi}arcsin(\rho_z)+C$.  Because $\rho_z=0$ implies $\rho_x=0$, thus, $C=0$, and the function relationship between $\rho_z$ and $\rho_x$ is:
\begin{equation}\label{sin}
\rho_z=sin\left(\frac{\pi}{2}\rho_x\right).
\end{equation}

\subsubsection{Case \uppercase\expandafter{\romannumeral6}}\label{CASE6}
For the case of $B(1,0.5)$ and $N(0,1)$, using  Eq.\eqref{Hermite}, Eq.\eqref{doublenormal2} and Eq.\eqref{1}, it has:
\begin{equation}
\begin{split}
\rho_x\frac{1}{2}=&\int_{0}^{+\infty}\int_{-\infty}^{+\infty} z_2\phi(z_1,z_2,\rho_z)dz_2dz_1\\
=&\sum_{k=0}^{\infty}\frac{\rho_z^k}{k!} \int_{0}^{+\infty}   \phi(z_1) H_k(z_1) dz_1\cdot\int_{-\infty}^{+\infty}z_2 H_k(z_2)\phi(z_2)dz_2\\
=&\sum_{k=0}^{\infty}\frac{\rho_z^k}{k!} \left(\sum_{k=0}^{}\frac{1}{2}+ \sum_{k=2n+1}^{}\frac{(-1)^n(2n)!}{\sqrt{2\pi}2^nn!} \right)\cdot\int_{-\infty}^{+\infty}z_2 H_k(z_2)\phi(z_2)dz_2\\
=&\frac{\rho_z}{\sqrt{2\pi}},
\end{split}
\end{equation}
then:
\begin{equation}
\begin{split}
\rho_x=\sqrt{\frac{2}{\pi}}\rho_z\leftrightarrow \rho_z=\sqrt{\frac{\pi}{2}}\rho_x.
\end{split}
\end{equation}

\subsubsection{Case \uppercase\expandafter{\romannumeral7}}\label{CASE7}
For the case of $B(1,0.5)$ and $lnN(\mu_2,\sigma_2^2)$, using  Eq.\eqref{doublenormal2},  Eq.\eqref{1} and   Eq.\eqref{c4}, it has:
\begin{equation}
\begin{split}
&\rho_x\frac{ \sqrt{e^{\sigma_2^2}-1}e^{\mu_2+\frac{\sigma_2^2}{2}}}{2}+\frac{e^{\mu_2+\frac{\sigma_2^2}{2}} }{2} =\int_{-\infty}^{+\infty}\int_{0}^{+\infty}
 e^{\mu_2+\sigma_2z_2}\phi(z_1,z_2,\rho_z)dz_1dz_2\\
=&e^{\mu_2+\frac{\sigma_2^2}{2}}\sum_{k_2=0}^{\infty}\sum_{k=0}^{\infty}\frac{\rho_z^k}{k!}\int_{0}^{+\infty}
H_{k}(z_1)\phi(z_1)dz_1\cdot \int_{-\infty}^{+\infty}
\frac{\sigma_2^{k_2}}{k_2!}H_{k_2}(z_2)H_{k}(z_2)\phi(z_2)dz_2\\
=&e^{\mu_2+\frac{\sigma_2^2}{2}}\sum_{k_2=0}^{\infty}\sum_{k=0}^{\infty}\frac{\rho_z^k}{k!}\left(\sum_{k=0}^{}\frac{1}{2}+ \sum_{k=2n+1}^{}\frac{(-1)^n(2n)!}{\sqrt{2\pi}2^nn!} \right)\cdot \int_{-\infty}^{+\infty}
\frac{\sigma_2^{k_2}}{k_2!}H_{k_2}(z_2)H_{k}(z_2)\phi(z_2)dz_2\\
=&e^{\mu_2+\frac{\sigma_2^2}{2}}\left(\frac{1}{2}+\sum_{n=0}^{\infty}\frac{(-1)^n(2n)!}{\sqrt{2\pi}2^nn!}\cdot\frac{\rho_z^{2n+1}}{(2n+1)!}\cdot\sigma_2^{2n+1} \right)\\
=&e^{\mu_2+\frac{\sigma_2^2}{2}}\Phi(\sigma_2\rho_z)\\
\end{split}
\end{equation}
The last step is due to Eq.\eqref{fff}.Then:
\begin{equation}\label{case7}
\begin{split}
\rho_x=\frac{2\Phi(\sigma_2\rho_z)-1}{\sqrt{e^{\sigma_2^2}-1}}.
\end{split}
\end{equation}

\subsubsection{Case \uppercase\expandafter{\romannumeral8}}\label{CASE8}
For the case of $N(0,1)$ and $lnN(\mu_2,\sigma_2^2)$, using  Eq.\eqref{doublenormal2} and Eq.\eqref{c4}, it has:
\begin{equation}
\begin{split}
\rho_x \sqrt{e^{\sigma_2^2}-1}e^{\mu_2+\frac{\sigma_2^2}{2}} =&\int_{-\infty}^{+\infty}\int_{-\infty}^{+\infty}
z_1e^{\mu_2+\sigma_2z_2}\phi(z_1,z_2,\rho_z)dz_1dz_2\\
=&e^{\mu_2+\frac{\sigma_2^2}{2}}\sum_{k_2=0}^{\infty}\sum_{k=0}^{\infty}\frac{\rho_z^k}{k!}\int_{-\infty}^{+\infty}
z_1H_{k}(z_1)\phi(z_1)dz_1\\
& \int_{-\infty}^{+\infty}
\frac{\sigma_2^{k_2}}{k_2!}H_{k_2}(z_2)H_{k}(z_2)\phi(z_2)dz_2\\
=&e^{\mu_2+\frac{\sigma_2^2}{2}}\sigma_2 \rho_z,\\
\end{split}
\end{equation}
then:
\begin{equation}
\rho_x=\frac{\sigma_2 \rho_z}{\sqrt{e^{\sigma_2^2}-1}}.
\end{equation}

\subsubsection{Case \uppercase\expandafter{\romannumeral9}}\label{CASE9}
For the case of $lnN(\mu_1,\sigma_1^2)$ and $lnN(\mu_2,\sigma_2^2)$, using  Eq.\eqref{doublenormal2} and Eq.\eqref{c4}, it has:
\begin{equation}
\begin{split}
&\rho_x\sqrt{(e^{\sigma_1^2}-1)(e^{\sigma_2^2}-1)}e^{\mu_1+\mu_2+\frac{\sigma_1^2}{2}+\frac{\sigma_2^2}{2}}+e^{\mu_1+\mu_2+\frac{\sigma_1^2}{2}+\frac{\sigma_2^2}{2}}=E[x_1x_2]\\
=&\int_{-\infty}^{+\infty}\int_{-\infty}^{+\infty}
e^{\mu_1+\sigma_1z_1}e^{\mu_2+\sigma_2z_2}\phi(z_1,z_2,\rho_z)dz_1dz_2.
\end{split}
\end{equation}

Using  Eq.\eqref{doublenormal2} and Eq.\eqref{c4}, it has:
\begin{equation}
\begin{split}
E[x_1x_2]
=&e^{\mu_1+\mu_2+\frac{\sigma_1^2}{2}+\frac{\sigma_2^2}{2}}\sum_{k_1=0}^{\infty}\sum_{k_2=0}^{\infty}\sum_{k=0}^{\infty}\frac{\rho_z^k}{k!}\int_{-\infty}^{+\infty}
\frac{\sigma_1^{k_1}}{k_1!}H_{k_1}(z_1)H_{k}(z_1)\phi(z_1)dz_1\cdot \\
 &\int_{-\infty}^{+\infty}
\frac{\sigma_2^{k_2}}{k_2!}H_{k_2}(z_2)H_{k}(z_2)\phi(z_2)dz_2\\
=&e^{\mu_1+\mu_2+\frac{\sigma_1^2}{2}+\frac{\sigma_2^2}{2}}\sum_{k=0}^{\infty}\frac{\sigma_1^{k}\sigma_2^{k}\rho_z^k}{k!}\\
=&e^{\mu_1+\mu_2+\frac{\sigma_1^2}{2}+\frac{\sigma_2^2}{2}}\cdot e^{\sigma_1\sigma_2\rho_z}.
\end{split}
\end{equation}
Then:
\begin{equation}\label{case9}
\begin{split}
&\rho_x\sqrt{(e^{\sigma_1^2}-1)(e^{\sigma_2^2}-1)}+1 =e^{\sigma_1\sigma_2\rho_z},\\
&\rho_x=\frac{e^{\sigma_1\sigma_2\rho_z}-1}{\sqrt{(e^{\sigma_1^2}-1)(e^{\sigma_2^2}-1)}}.
\end{split}
\end{equation}

\end{document}